\documentclass[12pt]{article}
\usepackage[table]{xcolor}

\usepackage{enumitem} 

\usepackage{amssymb, amsmath, amsthm}
\usepackage{graphicx,psfrag,epsf}
\usepackage{natbib}
\usepackage{multirow}
 \usepackage{subcaption}
\usepackage{graphicx}
\usepackage{hyperref}
\usepackage{times}
\usepackage{fullpage}
\usepackage{algorithm}
\usepackage{algpseudocode}
\usepackage{setspace}
\usepackage{textcomp}
\usepackage{booktabs}
\usepackage{xcolor}
\let\Algorithm\algorithm
\renewcommand\algorithm[1][]{\Algorithm[#1]\setstretch{0.8}}
\newcommand{\argmin}{\operatorname*{arg \ min}}

\theoremstyle{plain}

\newtheorem{prop}{Proposition}
\newtheorem{theorem}{Theorem}

\newtheorem{remark}{Remark}

\title{An explicit mean-covariance parameterization for\\
 multivariate response linear regression}

\author{Aaron J.\ Molstad$^{1}$\footnote{Correspondence: amolstad@ufl.edu}, 
 Guangwei Weng$^{2}$, Charles R.\ Doss$^{2}$, and Adam J.\ Rothman$^{2}$\\
$^{1}$Department of Statistics and Genetics Institute, University of Florida \\
$^{2}$School of Statistics, University of Minnesota}
\date{}
\begin{document}

\maketitle
\begin{abstract}
We develop a new method to fit the multivariate response linear regression model that exploits a parametric link between the regression coefficient matrix and the error covariance matrix.  Specifically, we assume that the correlations between entries in the multivariate error random vector are proportional to the cosines of the angles between their corresponding regression coefficient matrix columns, so as the angle between two regression coefficient matrix columns decreases, the correlation between the corresponding errors increases.  We highlight two models under which this parameterization arises: a latent variable reduced-rank regression model and the errors-in-variables regression model. We propose a novel non-convex weighted residual sum of squares criterion which exploits this parameterization and admits a new class of penalized estimators.  The optimization is solved with an accelerated proximal gradient descent algorithm. Our method is used to study the association between microRNA expression and cancer drug activity measured on the NCI-60 cell lines. An R package implementing our method, \texttt{MCMVR}, is available online.\end{abstract}


\doublespacing
\section{Introduction}
\label{s:intro}
Some regression analyses have more than one response
and these responses are typically associated.
When these responses are numerical variables, it is common to apply 
the multivariate response linear regression model. Let $y_i \in \mathbb{R}^q$ be the observed response for the $i$th subject, 
and let $x_i \in \mathbb{R}^p$ be the observed predictor for the $i$th subject. 
In the multivariate response linear regression model, $y_i$ is a realization of the random vector
\begin{equation} \label{eq:multivariate_model}
\textbf{Y}_i = \mu_* + \beta_*'x_i + \epsilon_i, \quad i=1, \dots, n, 
\end{equation}
where $\mu_*\in\mathbb{R}^q$ is the unknown intercept, 
$\beta_* \in \mathbb{R}^{p \times q}$ is the unknown regression coefficient matrix, 
and $\epsilon_1, \dots, \epsilon_n$ are 
independent copies of a mean zero random vector with covariance matrix $\Sigma_*$. Chapter 7 of 
\citet{pourahmadi2013high} gives a detailed overview of modern shrinkage methods 
that fit \eqref{eq:multivariate_model}. We review a subset of these methods here. 

Several shrinkage estimators of $\beta_{*}$ have been proposed through penalized least squares.
If the penalty separates across the columns of the optimization variable, then 
the estimate of $\beta_{*}$ can be computed with $q$ separate penalized least-squares regressions, 
e.g. lasso-penalized least squares.  Other penalized least-squares methods assume 
rows of $\beta_*$ are zero \citep{obozinski2011support, peng2010regularized}, assume $\beta_*$ is low rank \citep{izenman1975reduced,yuan2007dimension}, or assume both \citep{chen2012sparse}.

Under the additional assumption that the $\epsilon_i$'s are multivariate normal, \eqref{eq:multivariate_model} can be fit by minimizing a penalized negative Gaussian log-likelihood. These likelihood-based methods simultaneously estimate $\Sigma_*$ and $\beta_*$ \citep{izenman1975reduced, rothman2010sparse, yin2011sparse}. There also exist methods with two steps: they first estimate $\Sigma_*^{-1}$ and then plug this estimate into a penalized normal negative log-likelihood to estimate $\beta_*$ \citep{perrot2018variable}. There are also methods 
that add an assumption that the predictor and response are $(p+q)$-variate normal and develop estimators based on the inverse regression \citep{molstad2016indirect} or based on estimating the joint covariance matrix \citep{lee2012simultaneous}. 

We focus on methods that fit \eqref{eq:multivariate_model} by assuming that 
the error covariance matrix $\Sigma_*$ and the regression coefficient matrix $\beta_*$
are parametrically connected. One example is the envelope model, which assumes that the columns of $\beta_*$ are in a subspace spanned by eigenvectors of $\Sigma_*$ with small corresponding eigenvalues \citep{cook2015foundations}. Focusing on precision matrix estimation, \citet{pourahmadi1999joint} proposed a joint mean-covariance model based on an autoregressive interpretation of the Cholesky factor. 
In this manuscript, we consider a more explicit parametric connection between $\Sigma_*$ and $\beta_*$:
we propose
to fit \eqref{eq:multivariate_model} under the assumption that \begin{equation}\label{covariance_parameterization}
\Sigma_* \propto \beta_*'\beta_* + \sigma_{*\epsilon}^2 I_q,
\end{equation}
where $\sigma_{*\epsilon}^2\in (0,\infty)$ is unknown.
This parametrization links the angle between the $j$th and $k$th columns of $\beta_*$ and the correlation between the $j$th and $k$th responses; the motivation is that in some applications, two responses that depend on predictors in a similar way will also depend on unmeasured factors in similar ways. One setting in which this assumption may hold is the multivariate regression of cancer drug activity on microRNA expression profiles measured on the National Cancer Institute (NCI)-60 cell lines. In particular, because variation in cancer drug activity has been shown to be partly explained by -omic factors other than microRNA expression \citep{chen2017prediction}, it may be reasonable to assume that two drugs which depend on microRNA expression in a similar way also depend on unmeasured -omic factors (e.g., somatic mutations) in similar ways. In Section \ref{sec:data_analysis}, we show that assuming \eqref{covariance_parameterization} in this application leads to improved prediction accuracy and fitted models which may provide new biological insights about cancer drug activity.

Formally, \eqref{covariance_parameterization} implies that for each $(j,k) \in \{1,\ldots, q\}^2$, 
the cosine of the angle between the $j$th and $k$th column
of $\beta_{*}$ is proportional to the $(j,k)$th entry in $\Sigma_{*}$, 
so as the angle between the $j$th and $k$th column of $\beta_{*}$ decreases, 
the correlation between the corresponding errors increases.  
For example, if $\beta_{*j}'\beta_{*k} = 0$ (e.g., the $j$th and $k$th responses depend on different predictors), then it may 
be natural to assume that the $j$th and $k$th errors are uncorrelated since the $j$th
and $k$th responses 
relate to the $p$ predictors in distinct ways. 
If $\beta_{*j}'\beta_{*k}$ were instead relatively large, then 
it may  be natural to assume that the $j$th and $k$th errors are positively associated since the $j$th
and $k$th responses relate to the $p$ predictors in similar ways. 

\section{A new class of regression coefficient matrix estimators}
Let $\bar y = n^{-1} \sum_{i=1}^n y_i$ and $\bar x = n^{-1} \sum_{i=1}^n x_i$.  
Define $Y\in\mathbb{R}^{n \times q}$ to have $i$th row $(y_i - \bar y)'$
and define $X\in\mathbb{R}^{n \times p}$ to have $i$th row $(x_i - \bar x)'$.
Suppose that the  $\epsilon_i$'s are $q$-variate normal and 
$\Sigma_{*} = \sigma_{*1}^2 \beta_*'\beta_* + \sigma_{*2}^2 I_q $, where $\sigma_{*1}^2$ and $\sigma_{*2}^2$ are positive constants that represent the proportionality assumption in \eqref{covariance_parameterization}. Then two times the negative log likelihood (up to constants) evaluated at $(\beta, \sigma_{1}^2, \sigma_{2}^2)$
is
\begin{equation}\label{eq:log_likelihood_approach}
{\rm tr}\left\{ n^{-1}(Y - X\beta) (\sigma_1^2 \beta'\beta + \sigma_2^2 I_q)^{-1}(Y - X\beta)'\right\} + \log {\rm det}\left(\sigma_1^2 \beta'\beta + \sigma_2^2 I_q\right),
\end{equation}
where ${\rm tr}$ and ${\rm det}$ are the trace and determinant. A likelihood-based estimator of $\beta_*$ would require minimization over three optimization variables: $\beta$, $\sigma_{1}^2$, and $\sigma_{2}^2$. The scaling factor optimization variable $\sigma_1^2$ makes the function defined by \eqref{eq:log_likelihood_approach} difficult to minimize because it scales $\beta'\beta$ in both the trace and determinant terms. This optimization is made more difficult when one penalizes the entries of $\beta$. Moreover, since our goal is to estimate $\beta_*$, ideally, estimation of nuisance parameters $\sigma_{*1}^2$ and $\sigma_{*2}^2$ could be avoided entirely. Additional details about maximum likelihood estimation using \eqref{eq:log_likelihood_approach} are given in Section 4 of the Supplementary Material.

Instead, to estimate $\beta_*$ with the assumption in \eqref{covariance_parameterization}, 
we propose the class of estimators
\begin{equation}\label{non_convex_estimator}
\widehat{\beta}_{\tau, \lambda} = \argmin_{\beta \in \mathbb{R}^{p \times q}} \left\{ \mathcal{F}_{\tau}(\beta) + \frac{\lambda}{\tau} {\rm Pen}(\beta) \right\},
\end{equation}
where ${\rm Pen}(\cdot)$ is a user-specified penalty function;
$\tau$ and $\lambda$ are positive tuning parameters; and 
\begin{equation}\label{loss_function}
\mathcal{F}_{\tau}(\beta) =   {\rm tr}  \left\{ n^{-1} (Y - X\beta) ( \beta' \beta +  \tau I_q)^{-1}(Y - X\beta)' \right\}.
\end{equation}
The function defined in \eqref{loss_function} is similar to
\eqref{eq:log_likelihood_approach}, except that the scaling factor $\sigma_1^2$ and 
the log determinant term are removed.
This function generalizes the function proposed in \citet{gleser1973estimation}, which we discuss further in Section 3. 

We do not require a particular form for ${\rm Pen}(\cdot)$, but the 
algorithm we propose in Section 4 to solve \eqref{non_convex_estimator} will be most effective when the proximal operator of ${\rm Pen}(\cdot)$ can be computed efficiently.  In addition, a global minimizer for \eqref{non_convex_estimator} is only guaranteed to exist when ${\rm Pen}(\cdot)$ is coercive (see Remark 1 in Section 4.1). 

The function $\mathcal{F}_\tau$ is especially flexible due to the tuning parameter $\tau$. In particular, when $\tau \to \infty$, the matrix $\beta'\beta + \tau I_q$ becomes diagonally dominant and so $\mathcal{F}_\tau$ tends to the unweighted residual sum of squares. 
This has two main benefits: statistically, the theoretical properties of any penalized least squares estimator apply to \eqref{non_convex_estimator} by allowing $\tau \to \infty$ at a sufficient rate; practically, this gives practitioners the ability to determine whether, and to what extent, \eqref{covariance_parameterization} holds in a data-driven fashion. In particular, if \eqref{covariance_parameterization} does not hold, then our tuning parameter selection criterion, described in Section 1 of the Supplementary Material, should select $\tau$ sufficiently large so that $\widehat{\beta}_{\tau, \lambda}$ is effectively the same as the penalized least squares estimator. Viewed in this way, the tuning parameter $\tau$ is best thought of as analogous to the tuning parameter needed to specify the Huber loss function \citep{huber1964}. 

We do not intend that $\tau$ be interpreted as a ratio of unknown error variances. Were inference about $\beta_*$ or variance parameters in \eqref{eq:log_likelihood_approach} the goal of the practitioner, our method may not be appropriate since we treat the variances as nuisance parameters.




\section{Models connected to the parametric link}\label{sec:param_link}
\subsection{Errors-in-variables models}
The parameterization in \eqref{covariance_parameterization} holds under a multivariate response ``errors-in-variables'' linear regression model.  Consider the special case of \eqref{eq:multivariate_model} where for a latent (non-random) predictor $z_i \in \mathbb{R}^p$ for the $i$th subject, $y_i$ is assumed to be a realization of 
$$ \mathbf{Y}_i = \mu_* + \beta_*'z_i + \widetilde{\epsilon}_i,$$
where the $\widetilde{\epsilon}_i$'s are independent and identically distributed with ${\rm E}(\widetilde{\epsilon}_i) = 0,$ ${\rm Cov}(\widetilde{\epsilon}_i) = \gamma_*^2 I_q,$ for $i=1,\dots, n.$
Suppose we cannot measure $z_i$ exactly for $i=1, \dots, n$. Instead, we observe a realization of $\mathbf{X}_i = z_i + \mathbf{U}_i$ where the $\mathbf{U}_i$'s are independent and identically distributed with ${\rm E}(\mathbf{U}_i) = 0$ and ${\rm Cov}(\mathbf{U}_i) = \sigma_{*u}^2 I_p$ for $i = 1, \dots, n$; and $\mathbf{U}_i$ is independent of $\widetilde{\epsilon}_i$. It follows that
$$ \mathbf{Y}_i = \mu_* + \beta_*'z_i + \widetilde{\epsilon}_i  = \mu_* + \beta_*'\mathbf{X}_i - \beta_*'\mathbf{U}_i + \widetilde{\epsilon}_i.$$
Because we do not observe the realization of $\mathbf{U}_i$, 
$${\rm Cov}(\mathbf{Y}_i \mid \mathbf{X}_i = x_i) =  \beta_*' {\rm Cov}(\mathbf{U}_i) \beta_* + {\rm Cov}(\widetilde{\epsilon}) \propto \beta_*'\beta_* + \sigma_{*\epsilon}^2 I_q,$$
where $\sigma_{*\epsilon}^2 = \gamma_*^2/\sigma_{*u}^{2}.$ 

Fitting errors-in-variables models is a classical problem in low-dimensional
multivariate statistics. If one were willing to make distributional assumptions (e.g., normality) about the $\widetilde{\epsilon}_i$ and $\mathbf{U}_i$, then one could obtain maximum likelihood estimators by maximizing the joint likelihood for $(\mathbf{X}_1, \mathbf{Y}_1),\ldots, (\mathbf{X}_n, \mathbf{Y}_n)$ and treating the $z_i$'s as unknown parameters. Alternatively, one could fit the model that is conditional on the observed values of $\mathbf{X}_1,\ldots, \mathbf{X}_n$. \citet{gleser1973estimation} established an interesting connection between these two approaches in the low-dimensional case. In particular, \citet{gleser1973estimation} showed that in the special case where $\sigma_{*u}^2 = \gamma_*^2$ and $q = p$, the estimator obtained by maximizing the log-likelihood for the joint distribution of the predictor and response was equivalent to the estimate obtained by maximizing the weighted residual sum of squares:
\begin{equation}\label{eq:gleser_loss}
{\rm tr}\left\{ n^{-1}(Y - X \beta)(\beta'\beta  + I_q)^{-1} (Y - X \beta)'\right\},
\end{equation}
which is similar to the negative log-likelihood when $\mathbf{Y}_i \mid \mathbf{X}_i = x_i$ is multivariate normal. However, when $\sigma_{*u}^2$ and $\gamma_*^2$ are unequal, \eqref{eq:gleser_loss} may perform poorly.

Like \eqref{eq:gleser_loss}, our proposed weighted residual sum of squares criterion defined in \eqref{loss_function} is similar to the negative-log likelihood when one assumes $\epsilon_i$'s are multivariate normal. Unlike the function in \eqref{eq:gleser_loss}, our proposed criterion defined in \eqref{loss_function} replaces $\beta'\beta + I_q$ with $\beta'\beta + \tau I_q$. The introduction of the tuning parameter $\tau$ allows practitioners to account for the relationship between $\gamma_*^2$ and $\sigma_{*u}^2$ using cross-validation. 

Although the proposed criterion $\mathcal{F}_\tau$ is similar to the normal negative log-likelihood, it is not a valid likelihood function. However, we can justify its use based on the following result: 
\begin{theorem} 
Suppose $(Y, X)$ are generated from the multivariate response errors-in-variables model. If $\tau = \gamma_*^2/\sigma_{*u}^{2}$, then ${\rm E}\left\{\nabla \mathcal{F}_\tau (\beta_*) \right\} = 0$, that is, our estimator is Fisher consistent. 
\end{theorem}
We prove Theorem 1 in the Section 2 of the Supplementary Material. 
Note that we assume no particular distribution for the $\mathbf{U}_i$ or $\widetilde{\epsilon}_i$.

In Section 4 of the Supplementary Material, we compare estimates based on $\mathcal{F}_\tau$ to the maximum likelihood estimator (MLE) (based on \eqref{eq:log_likelihood_approach}) in low-dimensional settings. We show that with the tuning parameter $\tau$ chosen by cross-validation, the unpenalized version of \eqref{non_convex_estimator} performs similarly to the MLE under various data generating models. This provides some evidence that little efficiency is lost using $\mathcal{F}_\tau$ as an estimation criterion relative to the negative log-likelihood.

\subsection{Latent variable reduced-rank regression model}
Our parameterization also arises from a particular latent variable reduced-rank regression model \citep{velu2013multivariate}. This model assumes that the measured response for the $i$th subject is a realization of 
\begin{equation}\label{eq:latentRR_1}
\mathbf{Y}_i  = \mu_{*} + \mathcal{A}_* \mathbf{Z}_i  + \widetilde{\epsilon}_i, \quad (i=1, \dots, n),
\end{equation}
where $\mathcal{A}_* \in \mathbb{R}^{q \times r}$ with $r \leq {\rm min}(p,q)$, and $\widetilde{\epsilon}_1,\ldots, \widetilde{\epsilon}_n$ are independent and identically distributed with mean zero and covariance $\gamma_*^2 I_q$. In addition, 
\begin{equation}\label{eq:latentRR_2}
\mathbf{Z}_i = \mathcal{B}_*x_i + \mathbf{U}_i,  \quad (i=1, \dots, n),
\end{equation}
 where $\mathcal{B}_* \in \mathbb{R}^{r \times p}$ is a semiorthogonal matrix, the $x_i \in \mathbb{R}^{p}$ are the nonrandom values of the predictor for the $i$th subject, and $\mathbf{U}_1,\ldots, \mathbf{U}_n$ are independent and identically distributed 
with mean zero and covariance $\sigma_{*u}^2 I_r$.
It follows that
$${\rm E}(\mathbf{Y}_i)  = \mu_{*} + \mathcal{A}_*\mathcal{B}_*x_i,\quad 
{\rm Cov}(\mathbf{Y}_i)  =
\sigma_{*u}^2 \mathcal{A}_*\mathcal{A}_*' + \gamma_{*}^2 I_q, $$
so that the regression coefficient matrix is $\beta_* = \mathcal{B}_*'\mathcal{A}_*' \in \mathbb{R}^{p \times q}$ with ${\rm rank}(\beta_*) = r$. 
It is straightforward to verify that together, \eqref{eq:latentRR_1} and \eqref{eq:latentRR_2} imply the mean-covariance parameterization in \eqref{covariance_parameterization}.

\section{Computation}\label{sec:Computation}
Although $\mathcal{F}_\tau$ is not convex, it is differentiable and has Lipschitz continuous gradient over bounded sets. We formalize these properties in the following proposition. 
\begin{prop}
When $\tau > 0$, 
$$ \nabla \mathcal{F}_{\tau}(\beta) =  -2 n^{-1}\beta \Omega_\beta^{-1}(Y- X\beta)'(Y - X\beta)\Omega_\beta^{-1} - 2  n^{-1}X'Y \Omega_\beta^{-1} + 2 n^{-1}X'X \beta \Omega_\beta^{-1},$$
where $\Omega_\beta = \beta'\beta + \tau I_q.$ Moreover, $\nabla \mathcal{F}_{\tau}$ is Lipschitz over the set
$\mathcal{D}_\kappa = \left\{ \beta: \beta \in \mathbb{R}^{p \times q}, \|\beta\|_F \leq \kappa \right\}$ where $0 \leq \kappa < \infty,$
where $\|\cdot\|_F$ is the Frobenius norm. 
\end{prop}
We prove both parts of Proposition 1 in Section 2 of the Supplementary Material. 
\begin{remark}
Because \eqref{loss_function} is bounded below (since the trace of the product of two non-negative definite matrices is non-negative), as long as the penalty function is coercive, i.e., ${\rm Pen}(\beta) \to \infty$ as $\|\beta\|_F \to \infty,$ a global minimizer of \eqref{non_convex_estimator} over $\mathbb{R}^{p \times q}$ exists  and is in $\mathcal{D}_\kappa$ for some finite $\kappa$. 
\end{remark}

Given the properties established in Proposition 1 and Remark 1, we can use a proximal gradient descent algorithm to obtain a critical point of \eqref{non_convex_estimator} \citep{LiAcc2015}. Since $\mathcal{F}_{\tau}$ has a Lipschitz continuous gradient over the bounded set $\mathcal{D}_\kappa$, there exists a positive constant $L$ such that 
\begin{equation}\label{majorization}
 \mathcal{F}_{\tau}(\beta) \leq \mathcal{F}_{\tau}(\widetilde{\beta}) + {\rm tr}\left\{\nabla \mathcal{F}_{\tau}(\widetilde{\beta})'(\beta - \widetilde{\beta}) \right\} + \frac{L}{2}\|\beta - \widetilde{\beta}\|_F^2,
 \end{equation}
for all $\beta \in \mathcal{D}_\kappa$ and $\widetilde{\beta} \in \mathcal{D}_\kappa$. Thus, the right hand side of \eqref{majorization} is a \textit{majorizing function} of $\mathcal{F}_{\tau}$ at $\widetilde\beta$ (i.e., the right hand side of (9) is greater than or equal to $\mathcal{F}_\tau$ for all $\beta \in \mathcal{D}_\kappa$ with equality when $\beta = \widetilde\beta$). Hence, applying the majorize-minimize principle \citep{lange2016mm}, we use an algorithm whose iterates minimize the majorizing function at the previous iterate:
\begin{equation}\label{MM_update}
\beta^{(k+1)} = \argmin_{\beta \in \mathbb{R}^{p \times q}} \left\{ \mathcal{F}_{\tau}(\beta^{(k)}) + {\rm tr}\left\{\nabla \mathcal{F}_{\tau}(\beta^{(k)})'(\beta - \beta^{(k)}) \right\} + \frac{t_k}{2}\|\beta -\beta^{(k)}\|_F^2 + \frac{\lambda}{\tau} {\rm Pen}(\beta) \right\},
\end{equation}
where $t_k$ is a positive step-size parameter; and $\beta^{(k+1)}$ and  $\beta^{(k)}$ are the $(k+1)$th and $k$th iterates of the optimization variable corresponding to $\beta$, respectively. This way, for sufficiently large $t_k$, we are guaranteed that $\mathcal{F}_\tau(\beta^{(k+1)}) + \frac{\lambda}{\tau} {\rm Pen}(\beta^{(k+1)}) \leq \mathcal{F}_\tau(\beta^{(k)}) + \frac{\lambda}{\tau} {\rm Pen}(\beta^{(k)})$ for all $k$. 

The iterate in \eqref{MM_update} can be written in the more familiar notation: 
$$ \beta^{(k+1)} = {\rm Prox}_{t_k^{-1}\frac{\lambda}{\tau}  {\rm Pen}}\left\{\beta^{(k)} -  t_k^{-1} \nabla \mathcal{F}_\tau(\beta^{(k)}) \right\},$$
where, using the notation from \citet{parikh2014proximal}, ${\rm Prox}_f$ denotes the proximal operator of the function $f$: 
$$ {\rm Prox}_{f}(y) = \argmin_{x}\left\{ \frac{1}{2}\|x - y\|_F^2 + f(x)\right\}.$$
The proximal operator can be computed efficiently for a broad class of convex and non-convex penalty functions. In Table 1 of the Supplementary Material, we provide closed form solutions of four proximal operators corresponding to convex penalties used in multivariate response linear regression. For example, if one used the $L_1$ norm as a penalty, the proximal operator is simply the soft-thresholding operator. 

With an appropriate choice of step size parameter $t_k$ for each $k$, iterates generated from \eqref{MM_update} are guaranteed to monotonically decrease the objective function value. However, this is not sufficient to ensure that the iterates converge to a critical point. In our implementation, we use an accelerated variation of the proximal gradient descent algorithm proposed by \citet{LiAcc2015} specifically designed for solving non-convex optimization problems which ensures that iterates converge to a critical point of \eqref{non_convex_estimator}. 


The complete algorithm is sketched in Algorithm 1. We implement this algorithm, along with a number of auxiliary functions, in the the R package \texttt{MCMVR}, which is available for download at \href{http://github.com/ajmolstad/MCMVR}{github.com/ajmolstad/MCMVR}. 
 \begin{algorithm}
\begin{enumerate}
\item[]\hspace{-20pt}\textbf{Algorithm 1:} Initialize $\beta^{(0)} = \beta^{(-1)} = \widetilde{\beta}^{(0)} = \overline{\beta}^{(0)}$, $\alpha^{(0)} = \alpha^{(-1)} = 1$, and set $k=0$. 
\
	\item Compute $\widetilde{\beta}^{(k)} \leftarrow \beta^{(k)} + \frac{\alpha^{(k-1)}}{\alpha^{(k)}}\left(\overline{\beta}^{(k)} - \beta^{(k)}\right) + \left(\frac{\alpha^{(k-1)} - 1}{\alpha^{(k)}}\right)\left(\beta^{(k)} - \beta^{(k-1)}\right).$
	\item Compute $\overline{\beta}^{(k+1)} \leftarrow {\rm Prox}_{ \overline{t}_k^{-1}\frac{\lambda}{\tau} {\rm Pen}} \left\{\widetilde{\beta}^{(k)} - \overline{t}_k^{-1} \nabla_\beta \mathcal{F}_{\tau}(\widetilde{\beta}^{(k)})  \right\}.$
	\item Compute $\Gamma^{(k+1)} \leftarrow {\rm Prox}_{ t_k^{-1}\frac{\lambda}{\tau} {\rm Pen}} \left\{\beta^{(k)} - t_k^{-1} \nabla_\beta \mathcal{F}_{\tau}(\beta^{(k)})  \right\}.$
	\item Set $\alpha^{(k+1)} \leftarrow (1 + \sqrt{1 + 4\alpha^{2(k)}})/2.$ 
 	\item Set $\beta^{(k+1)} \leftarrow \left\{ \begin{array}{c l}
\overline{\beta}^{(k+1)} :& \mathcal{F}_{\tau}(\overline{\beta}^{(k+1)})  + \frac{\lambda}{\tau}{\rm Pen}(\overline{\beta}^{(k+1)}) < \mathcal{F}_{\tau}(\Gamma^{(k+1)}) + \frac{\lambda}{\tau}{\rm Pen}(\Gamma^{(k+1)})\\
\Gamma^{(k+1)} :& \text{otherwise}
\end{array} \right.$
\item Set $k \leftarrow k + 1$, and return to Step 1. 
\end{enumerate}
\end{algorithm}

Step sizes $t_k$ and $\overline{t}_k$ from Algorithm 1 are chosen using backtracking line searches for both Step 2 and Step 3 of Algorithm 1. See Algorithm 2 of the Supplementary Material to \citet{LiAcc2015} for the exact version of the algorithm we implement. An application of Theorem 1 of \citet{LiAcc2015} ensures that the iterates generated by our algorithm are bounded and that the sequence of iterates converge to a critical point of \eqref{non_convex_estimator}. In Section B of the Supplementary Material, we describe how to select tuning parameters by cross-validation, and how to determine a reasonable set of candidate tuning parameter values.

\section{Generalizations}\label{sec:Generalizations}
In Section 3 of the Supplementary Material, we discuss an extension of our method to settings where only a subset of the response vectors are observed. In the remainder of this section, we describe how to modify our method to deal with covariates unrelated to $\Sigma_*$ and how to standardize predictors for model fitting. 
 
\subsection{Covariates unrelated to $\Sigma_*$}
An important generalization of our estimator includes a set of measured covariates $v_i \in \mathbb{R}^k$ such that
$$ \mathbf{Y}_i = \mu_* + \beta_*'x_i + \eta_*' v_i + \epsilon_i, \quad (i=1,\dots, n), $$
and the unknown coefficient matrix $\eta_* \in \mathbb{R}^{k \times q}$ is not parametrically related to $\Sigma_* \propto \beta_*'\beta_* + \sigma_{*\epsilon}^2 I_q$. When our method is motivated through the errors-in-variables model, this may occur when some covariates or confounders are measured without error, e.g., $x_i$ is some -omic profile measured with error and $v_i$ are clinical/demographic variables.

Let $\bar{v} = n^{-1}\sum_{i=1}^n v_i$. Define $V \in \mathbb{R}^{n \times k}$ to have $i$th row $(v_i - \bar{v})'$ and suppose $V$ has full rank. For this scenario, we propose the class of penalized estimators: 
$$
(\tilde{\beta}_{\tau, \lambda}, \tilde{\eta}_{\tau, \lambda}) = \argmin_{\beta \in \mathbb{R}^{p \times q}, \eta \in \mathbb{R}^{k \times q}} \left[ {\rm tr}\left\{n^{-1} (Y - V\eta - X\beta)'(Y - V\eta - X\beta) [\beta'\beta + \tau I_q]^{-1}\right\} + \frac{\lambda}{\tau}{\rm Pen}(\beta) \right].
$$
Using the first order conditions for $\eta$ and letting 
${\rm P}^\perp_{V} = I_n - V(V'V)^{-1}V'$, we replace $Y$ with $\tilde{Y} = {\rm P}^\perp_{V} Y$, $X$ with $\tilde{X} = {\rm P}^\perp_{V} X$, and solve a modified version of our estimator:
$$ \tilde{\beta}_{\tau, \lambda} =  \argmin_{\beta \in \mathbb{R}^{p \times q}} \left[ {\rm tr}\left\{n^{-1} (\tilde{Y}  - \tilde{X} \beta)'(\tilde{Y} - \tilde{X}\beta)[\beta'\beta + \tau I_q]^{-1}\right\} + \frac{\lambda}{\tau}{\rm Pen}(\beta) \right],$$
so that $\tilde{\eta}_{\tau, \lambda} = (V'V)^{-1}V'(Y - X\tilde{\beta}_{\tau, \lambda})$. Thus computing our estimator with the additional covariates can be done immediately from Algorithm 1 and $(\tilde{\beta}_{\tau, \lambda}, \tilde{\eta}_{\tau, \lambda})$ will satisfy the first order conditions for $(\tilde{\beta}_{\tau, \lambda}, \tilde{\eta}_{\tau, \lambda})$.

\subsection{Predictor standardization and dependent measurement errors}
An important property in regression coefficient matrix estimation is invariance under changes in scale of the predictor. Of course, our method is not invariant as the scale of the predictors affects the magnitude of entries in $\beta_*$, which affects the weight in the weighted residual sum of squares criterion we propose. However, we can easily generalize our estimator to allow for standardization, and in the context of the errors-in-variables model, allow for dependent measurement errors (assuming their covariance were known). 

The generalized version of $\mathcal{F}_{\tau}$ we propose is 
$$ \mathcal{H}_{\tau, \Phi}(\beta) = \left\{ n^{-1} (Y - X\beta)'(Y - X\beta) \left[ \beta'\Phi\beta +  \tau I_q\right]^{-1} \right\},$$
where $\Phi \in \mathbb{R}^{p \times p}$ is some user-specified, symmetric and nonnegative definite weight matrix. 
Notice, were one to standardize predictors so that columns of $\tilde{X}$ had columnwise average zero and unit standard deviation, we could write $X \beta = \tilde{X}\tilde{\beta}$ where $\tilde{\beta} = S^{-1} \beta$ and $S \in \mathbb{R}^{p \times p}$ has the inverse standard deviations of the predictors on its diagonal and zeros elsewhere. Thus, if $\Phi = S'S$, it follows that  
$${\rm tr}\left\{ n^{-1} (Y - \tilde{X}\tilde{\beta})'(Y - \tilde{X}\tilde{\beta}) [ \tilde{\beta}'S'S\tilde{\beta} +  \tau I_q ]^{-1} \right\} = \mathcal{F}_\tau(\beta),$$
where $\beta = S \tilde{\beta}$. 

If the model from Section 3.1 were assumed to hold, then this generalization could also be used in the case that measurement errors (i.e., the $\mathbf{U}_i$ from Section 3.2) have covariance $\Sigma_{*u}$, which is known or can be estimated reliably from external data. In this case, it would follow that 
${\rm Cov}(\mathbf{Y}_i \mid \mathbf{X}_i = x_i) = \beta_*'\Sigma_{*u} \beta_* + \gamma_{*}^2 I_q,$
so that a more appropriate weighted residual sum of squares would be $\mathcal{H}_{\tau, \Sigma_{*u}}$. 

The same computational approach developed in Section 4 of the main manuscript can be used since $\mathcal{H}_{\tau, \Phi}$ is differentiable and has Lipschitz continuous gradient over $\mathcal{D}_\kappa$.

\section{Simulation studies}
\subsection{Data generating models and performance metrics}\label{subsec:dataGen}
We compare the performance of our method to relevant competitors under three distinct data generating models. Under the first two models, when conditioning on the observed predictors, the mean and covariance of the response are the same in both models. However, the two models differ in a fundamental way:  in the first model, we observe a corrupted version of the ``true'' predictor so that conditioning on the observed predictor, the model in \eqref{eq:multivariate_model} and \eqref{covariance_parameterization} holds.  In the second model, we observe the ``true'' predictor and the covariance has the parameterization in \eqref{covariance_parameterization}. In the third model we consider, there are ``errors-in-variables'' but the covariance parameterization \eqref{covariance_parameterization} does not hold.

In the following, for one hundred independent replications with $p=200$ and $q = 50$, we generate $n=100$ independent copies of $(\mathbf{Y}, \mathbf{X})$.\\\\
-- \textbf{Model 1:} We first generate $n$ independent copies of $\mathbf{Z} \sim {\rm N}_p(0, \Sigma_{*Z})$ where the $(j,k)$th entry of $\Sigma_{*Z}$ equals $0.5^{|j-k|}$. Then, conditional on $\mathbf{Z} = z$, we generate a realization of $\mathbf{X}$ and $\mathbf{Y}$, 
$$ \mathbf{Y} = \beta_*' z + \epsilon, \quad \mathbf{X} = z + \mathbf{U}, $$
where $\mathbf{U} \sim {\rm N}_p(0, \sigma_{*u}^2 I_p)$ and $\epsilon \sim {\rm N}_p(0, \gamma_{*}^2 I_q)$ so that 
${\rm E}(\mathbf{Y} \mid \mathbf{X}=x) = \beta_*'x$ and ${\rm Cov}(\mathbf{Y} \mid \mathbf{X} = x) = \sigma_{*u}^2 \beta_*'\beta_* + \gamma_{*}^2 I_q,$
 with $\gamma_{*}^2 = 3$ and $\sigma_{*u}^2$ varying across settings.\medskip

\noindent
-- \textbf{Model 2:}  We first generate $n$ independent copies of $\mathbf{X} \sim {\rm N}_p(0, \Sigma_{*X})$ where the $(j,k)$th entry of $\Sigma_{*X}$ equals $0.7^{|j-k|}$. Then, conditional on $\mathbf{X} = x,$ we generate a realization of
\begin{equation}\label{eq:mean_covariance_dgm} 
\mathbf{Y} = \beta_*' x + \epsilon,
\end{equation}
where $\epsilon \sim {\rm N}_q(0, \sigma_{*u}^2 \beta_*'\beta_* + \gamma_{*}^2 I_q)$ with $\gamma_{*}^2 = 3$ and $\sigma_{*u}^2$ varying.  Note that Model 2 differs from Model 1 as under Model 1, the covariance of the measured predictors $\mathbf{X}$ is $\Sigma_{*X} = \Sigma_{*Z} + \sigma^2_{*u}I_p.$  \medskip

\noindent
-- \textbf{Model 3:} We first generate data in the same manner as Model 1 except
   where $\mathbf{U} \sim {\rm N}_p(0, \sigma_{*u}^2 I_p)$ and $\epsilon \sim {\rm N}_p(0, \gamma_*^2 \Sigma_{*E})$ where $[\Sigma_{*E}]_{j,k} = 0.7 \cdot \mathbf{1}(j \neq k) + \mathbf{1}(j = k)$,  where $\sigma_{*u}^2 = 0.50$,
   and where $\gamma_*^2$ is varying  across settings.

To generate $\beta_*$, we randomly construct three active sets of three variables each: let $a_k = \{a_{k,1}, a_{k,2}, a_{k,3}\} \subset \{1,\dots, p \}$ for $k=1, 2, 3$ with $\cap_{k=1}^3 a_{k}$ being empty. Then, for $l=1, \dots, q$, we randomly choose $k \in \left\{ 1, 2, 3 \right\}$ with probability $1/3$ each, and set either $[\beta_{*}]_{ (a_{k,j}), l} = -2$ or $[\beta_{*}]_{(a_{k,j}), l} = 2$, with equal probability for $j=1, \dots, 3$. We also select three additional elements of $[\beta_{*}]_{\cdot, l}$ to be $-1$ or $1$. That is, each column of $\beta_*$ has six nonzero entries: three entries have magnitudes $2$ and three entries have magnitudes $1$.  Under this construction, $\beta_*'\beta_*$ is approximately block diagonal with three blocks of similar size. 

We consider multiple performance metrics. The first we consider is \textit{model error} \citep{breiman1997predicting} for the observed predictor:
$\|\Sigma_{*X}^{1/2}(\widehat{\beta} - \beta_*)\|_F^2.$
Following \citet{datta2017cocolasso}, when data are generated from Model 2, we also measure the \textit{latent model error}, i.e, model error under the unobserved predictor $Z$: 
$\|\Sigma_{*Z}^{1/2}(\widehat{\beta} - \beta_*)\|_F^2.$
Latent model error would be relevant if the true predictor may be observed in future studies. 
In addition, for both models we also measure (squared) Frobenius norm error: 
$\|\widehat{\beta} - \beta_*\|_F^2,$
and out-of-sample prediction error: 
$\|Y_T - X_T \widehat{\beta}\|_F^2/q n_T.$
To compute the out-of-sample prediction error, in each replication we generate an independent test set of size $n_T = 1000$, where $Y_T \in \mathbb{R}^{n_T \times q}$ and $X_T \in \mathbb{R}^{n_T \times p}$, using the same data generating model as in the training data. It is important to note that out-of-sample prediction error and model error are distinct metrics. Model error measures how well an estimator predicts the mean function, whereas prediction error measures sum of squared residuals on a testing set. We also measure true and false positive identification of nonzero entries in $\beta_*$ to assess variable selection accuracy of the methods. 

\subsection{Competing methods}
We compare our method to two versions of the method proposed by \citet{datta2017cocolasso}, two versions of the $L_1$-penalized least squares estimator, and a two-step convex approximation to \eqref{non_convex_estimator}. Throughout, let $|A|_1 = \sum_{j,k}|A_{j,k}|$ for a matrix or vector $A$, and let $A_{\cdot, j}$ denote the $j$th column of $A$. 

For the case that $q=1$ and data are generated from an errors-in-variables model (e.g., Model 2), \citet{datta2017cocolasso} proposed the convex-conditioned lasso estimator. Their estimator can be naturally extended to the multivariate setting: they replace $X$ and $Y$ in the least squares criterion with versions adjusted to account for the measurement error. Assuming that $\sigma_{*u}^2$ were known, the estimator of \citet{datta2017cocolasso} modifies the unbiased sample covariance matrix: 
\begin{equation}\label{S_tilde}
 \widetilde{\Sigma} = \argmin_{S_1\geq 0} \|\widetilde{S} - S_1\|_{\max}, \quad \text{where } \widetilde{S} = n^{-1}X'X - \sigma_{*u}^2 I_p.
\end{equation}
Then the multivariate response generalization of their estimator is 
\begin{equation}\label{cc_lasso}
 \argmin_{\beta \in \mathbb{R}^{p \times q}} \left\{ {\rm tr}\left( \beta'\widetilde{\Sigma} \beta - 2\beta'\rho \right) + \lambda {\rm Pen}(\beta) \right\},
\end{equation}
where $\rho = n^{-1} X'Y$, which can be solved using penalized least squares.
When $\lambda {\rm Pen}(\beta)$ is replaced by  $ \sum_{j=1}^q \lambda_j {\rm Pen}(\beta_{\cdot, j})$, the estimator in \eqref{cc_lasso} is equivalent to performing $q$ separate convex-conditioned lasso estimation problems. The estimator in \eqref{cc_lasso} would not be equivalent to $q$ separate estimators if only one tuning parameter were used for all $q$ regressions, or any of the penalties from Table 1 of the Supplementary Material other than the $L_1$ norm was used.  We now formally state the competitors we consider:

\noindent -- \texttt{CoCo-1}: The estimator defined in \eqref{cc_lasso} with $\lambda {\rm Pen}(\beta) = \lambda |\beta|_1$ and $\lambda$ chosen by five-fold cross-validation, using the modified cross-validation procedure (averaged over the $q$ responses) proposed in \citet{datta2017cocolasso}. We treat the value of $\sigma_{*u}^2$ as known. \medskip

\noindent -- \texttt{CoCo-q}: The estimator defined in \eqref{cc_lasso} with $\lambda {\rm Pen}(\beta)$ replaced by $\sum_{j=1}^q \lambda_j |\beta_{\cdot, j}|_1$ and the $\lambda_j$ each chosen by five-fold cross-validation, using the modified cross-validation procedure proposed in \citet{datta2017cocolasso} for $j=1,\dots,q$. We treat the value of $\sigma_{*u}^2$ as known. \medskip

\noindent 
-- \texttt{CV-CoCo-q},  \texttt{CV-CoCo-1}: The same estimators as \texttt{CoCo-q} and \texttt{CoCo-1}, except $\sigma_{*u}^2$ is unknown and treated as a tuning parameter. We select both $\lambda$ and the $\sigma_{*u}^2$ value by five-fold cross-validation, using the modified cross-validation procedure proposed in \citet{datta2017cocolasso}.\medskip

\noindent 
-- \texttt{Lasso-q}: The $L_1$-penalized least squares estimator 
\begin{equation}\label{eq:lasso_q} \argmin_{\beta \in \mathbb{R}^{p \times q}} \left\{\frac{1}{n}\|Y - X \beta\|_F^2 + \sum_{j=1}^q \lambda_j |\beta_{\cdot, j}|\right\}
\end{equation}
within tuning parameters $\lambda_j$ chosen to minimize prediction error in five-fold cross-validation for $j=1,\dots,q$ separately.\medskip

\noindent 
-- \texttt{Lasso-1}: The estimator defined in \eqref{eq:lasso_q} except the tuning parameter $\lambda_j = \lambda$ for $j=1,\dots,q$ with $\lambda$ chosen to minimize prediction error averaged over the $q$ responses in five-fold cross-validation.\medskip

\noindent 
-- \texttt{MC}: The version of our estimator \eqref{non_convex_estimator} with ${\rm Pen}(\beta) =|\beta|_1,$ with tuning parameters $\lambda$ and $\tau$ chosen using the five fold cross-validation procedure described in Section 1 of the Supplementary Material.\medskip

\noindent 
The sixth competitor we consider, \texttt{CA}, is a two-step convex approximation to \eqref{non_convex_estimator}. Given a initial estimator $\widetilde{\beta}$, we re-estimate $\beta_*$ using 
\begin{equation} \label{convex_approx}
\overline{\beta} = \argmin_{\beta \in \mathbb{R}^{p \times q}} \left[ {\rm tr}\left\{n^{-1}(Y - X\beta)(\widetilde{\beta}' \widetilde{\beta} + \tau I_q)^{-1}(Y - X \beta)'\right\}  + \frac{\lambda}{\tau} |\beta|_1 \right].
\end{equation}
The estimator defined in \eqref{convex_approx} is computed using the coordinate descent algorithm of \citet{rothman2010sparse}. 

In our implementation of the estimator \texttt{CA}, we obtain $\widetilde{\beta}$ using \texttt{Lasso-q} and select the tuning parameters $\tau$ and $\lambda$ to minimize prediction error in five-fold cross-validation. The estimator \texttt{CA} is included to help illustrate the importance of simultaneous estimation of the covariance matrix and regression coefficients under \eqref{covariance_parameterization}. 

\begin{figure}[t!]
\centerline{
\hfill\includegraphics[width=6cm]{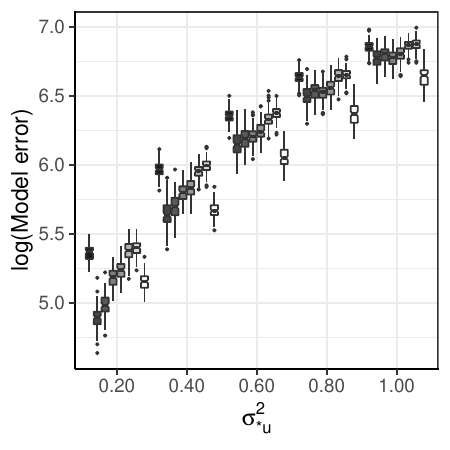}
\includegraphics[width=6cm]{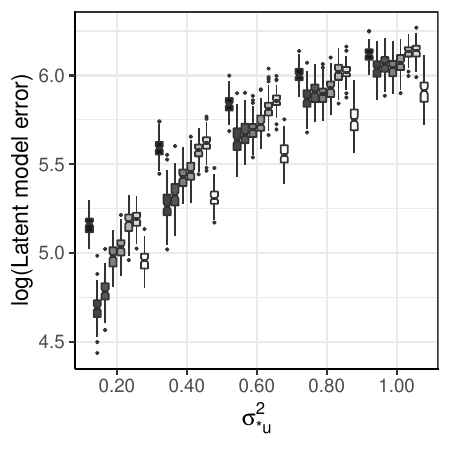}\hfill\includegraphics[width=6cm]{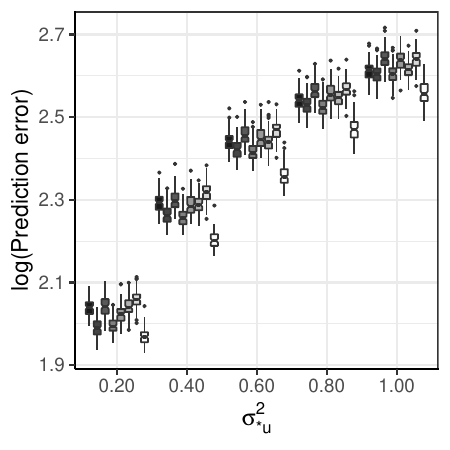}\hfill}
\centerline{\hfill\includegraphics[width=15cm]{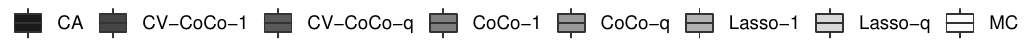}\hfill}


\caption{Log model error, log latent model error, and log prediction error for the eight candidate methods over one hundred independent replications under Model 1.}\label{fig:Model1_Results}
\end{figure}

\subsection{Results}
\begin{figure}
\centerline{
\hfill\includegraphics[width=6cm]{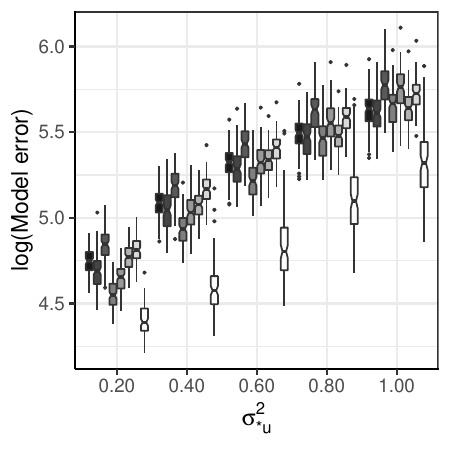}
\includegraphics[width=6cm]{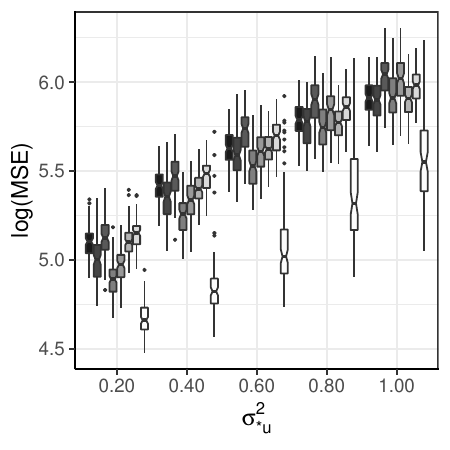}\hfill\includegraphics[width=6cm]{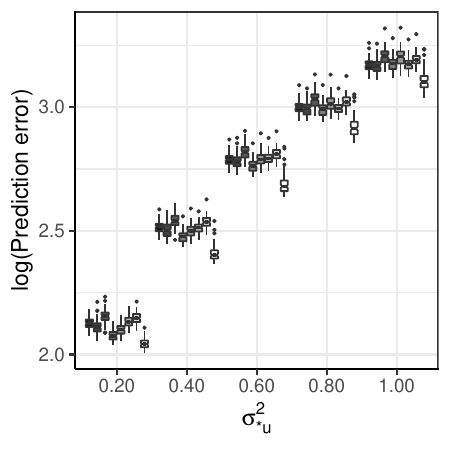}\hfill}
\centerline{\hfill\includegraphics[width=15cm]{Plots/Legend2.pdf}\hfill}


\caption{Log model error, log mean squared error, and log prediction error for the eight candidate methods over one hundred independent replications under Model 2.}\label{fig:Model2_Results}
\end{figure}

Results for Models 1-3 are displayed in Figures 1-3, respectively. We first discuss results under Model 2, which are displayed in Figure \ref{fig:Model2_Results}.
As $\sigma_{*u}^2$ increases for Model 2 we see that our proposed method, \texttt{MC}, outperforms all competitors in terms of model error, Frobenius norm error, and prediction error. Amongst the competitors, \texttt{CoCo-1} is best when $\sigma_{*u}^2 < 1$. When $\sigma_{*u}^2 = 1$, \texttt{CoCo-1} performs similarly to \texttt{Lasso-1} and the convex approximation of our method \texttt{CA}. Interestingly, we see both \texttt{Lasso-1} and \texttt{CoCo-1} outperform their counterparts which select tuning parameters separately for each response. A similar result was observed in the simulations of \citet{molstad2016indirect}. 

In Figure \ref{fig:Model1_Results}, we display results for Model 1. Unlike in Model 2, however, we see that \texttt{CV-CoCo-1} performs better than all other competitors even as $\sigma_{*u}^2$ becomes large. Our method again outperforms all competitors when $\sigma_{*u}^2$ is greater than 0.25. This is particularly notable for the latent model error, where the \texttt{CoCo} variants outperform both \texttt{Lasso} variants.

For Models 1 and 2, although not displayed, we also considered the case that $\sigma_{*u}^2 = 0$, i.e., \eqref{covariance_parameterization} does not hold. When $\sigma_{*u}^2 = 0$ our method performed similarly to \texttt{Lasso-1}, as did the method of \citet{datta2017cocolasso}. This result illustrates the property of \eqref{loss_function} highlighted in Section 3: when \eqref{covariance_parameterization} does not hold, cross-validation should select a $\tau$ large enough so that \eqref{loss_function} is effectively least squares.

\begin{figure}[t]
\centerline{
\hfill\includegraphics[width=6cm]{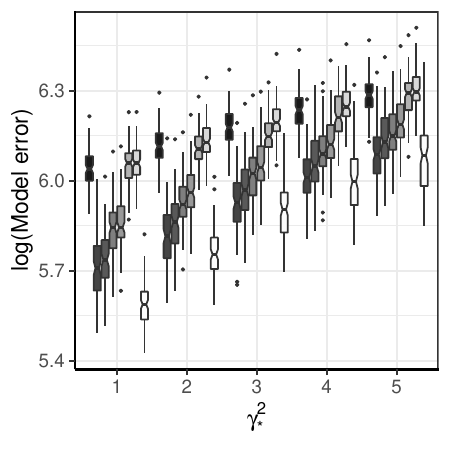}
\includegraphics[width=6cm]{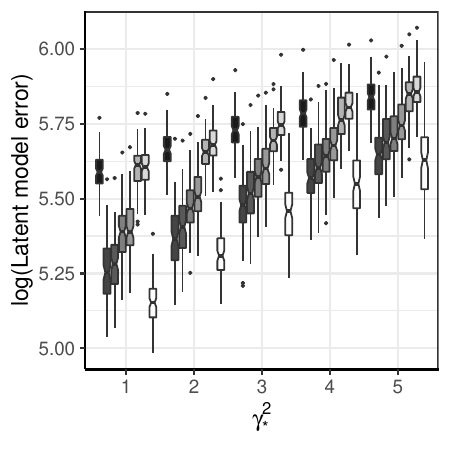}\hfill\includegraphics[width=6cm]{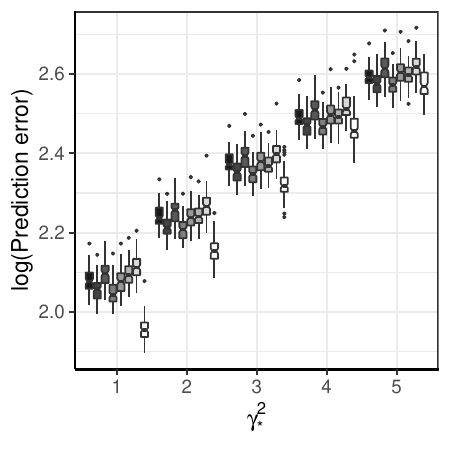}\hfill}
\centerline{\hfill\includegraphics[width=15cm]{Plots/Legend2.pdf}\hfill}


\caption{Log model error, log latent model error, and log prediction error for the eight candidate methods over one hundred independent replications under Model 3.}\label{fig:Model3_Results}
\end{figure}

In Figure \ref{fig:Model3_Results}, we display model error, latent model error, and prediction error results under Model 3. Recall that under Model 3, \eqref{covariance_parameterization} does not hold: error correlations are not entirely determined by $\beta_*$. Nevertheless, we see that for all the values of $\gamma_*^2$ which we consider, \texttt{MC} performs best amongst the competing methods. As $\gamma_*^2$ becomes larger, the distinction between methods becomes smaller. This is intuitive given that a large $\gamma_*^2$ corresponds to a smaller signal to noise ratio. 

Model selection results are displayed in Table 2 and 3 of the Supplementary Material. We see that as $\sigma_{*u}^2$ increases under Model 1 and 2, the true positive rate of our method, \texttt{MC}, tends to be significantly higher than any of the competing methods. Interestingly, both \texttt{CoCo-1} and \texttt{CoCo-q} have smaller false positive rates than \texttt{MC} when $\sigma_{*u}^2 \geq 0.50$, but both have significantly smaller true positive rates than \texttt{MC}. This may partly explain the difference in performance between the \texttt{CoCo} methods and \texttt{MC}. Results are similar under Model 3.


In Section 5 of the Supplementary Material, we present results from an additional simulation study under the latent reduced-rank regression model discussed in Section 3.2. We compare two ridge regression variants, nuclear norm-penalized least squares, a nuclear norm-penalized variation of \texttt{CA}, and our proposed estimator with ${\rm Pen}(\beta)$ equal to the nuclear norm of $\beta$. Results are similar to those under Model 1 and Model 2: our proposed estimator \eqref{non_convex_estimator} outperforms the penalized least squares variants in almost every replication when $\sigma_{*u}^2 > 0$. In Section 7 of the Supplementary Material, we also present simulation study results under a version of Model 3 with $[\Sigma_{*E}]_{j,k} = 0.9^{|j-k|}$.


In Section 6 of the Supplementary Material, we consider an additional competitor, \texttt{MC-Or}, which is the estimator \eqref{non_convex_estimator} with $\tau = \gamma_*^2/\sigma_{*u}^2$ and $\lambda$ selected by cross-validation. As was observed with the \texttt{CV-CoCo} variants, this version of our method perform substantially worse that that which treats $\tau$ as a tuning parameter. 



\section{Cancer drug activity data analysis}\label{sec:data_analysis}
In this section, we use our method to analyze a dataset consisting of microRNA expression profiles and cancer drug activity measurements on the NCI-60 cell lines \citep{shoemaker2006nci60}. The NCI-60 cell lines are a panel of 60 human tumor cancer cell lines representing leukemia, melanoma, and numerous cancers coming from distinct tissue types: breast, central nervous system, colon, lung, prostate, ovary, and kidney.

Modelling the relationship between -omic profiles and cancer drug activity is a topic of recent interest: see \citet{chen2017prediction} and references therein. The interaction between microRNA expression (predictors) and cancer drug activity (response) is especially relevant given that microRNAs are believed to play a key role in the development of many cancers as they can act as tumor suppressors or oncogenes \citep{peng2016role}. Previous studies have found significant correlations between microRNA expression and activity in certain cancer drugs in these particular cell lines \citep{liu2010mrna}. 

The particular dataset we analyze is publicly available through the \texttt{FRCC} R package on CRAN \citep{cruz2014fast}. These particular data were originally obtained from the CellMiner Database via \href{http://discover.nci.nih.gov/cellminer}{http://discover.nci.nih.gov/cellminer}). Following the analysis of \citet{cruz2014fast}, we restrict our attention to $q=15$ drugs (specifically, Topoisomerase II Inhibitors) belonging to the A118 drug dataset (http://dtp.cancer.gov). See Table 3 of \citet{cruz2014fast} for more information about the particular drugs in the A118 dataset. The microRNA expression profiles were measured on a Agilent Human microRNA Microarray and consist of $p=365$ microRNAs which had sufficient expression in at least 10\% of cell lines \citep{liu2010mrna}. According to the CellMiner Database, drug activity levels are defined as 50\% growth inhibition (molar concentration), and microRNA expression levels are on the log-base-2 scale. 

\begin{figure}[!tp]
\centerline{
\hfill\includegraphics[width=7cm]{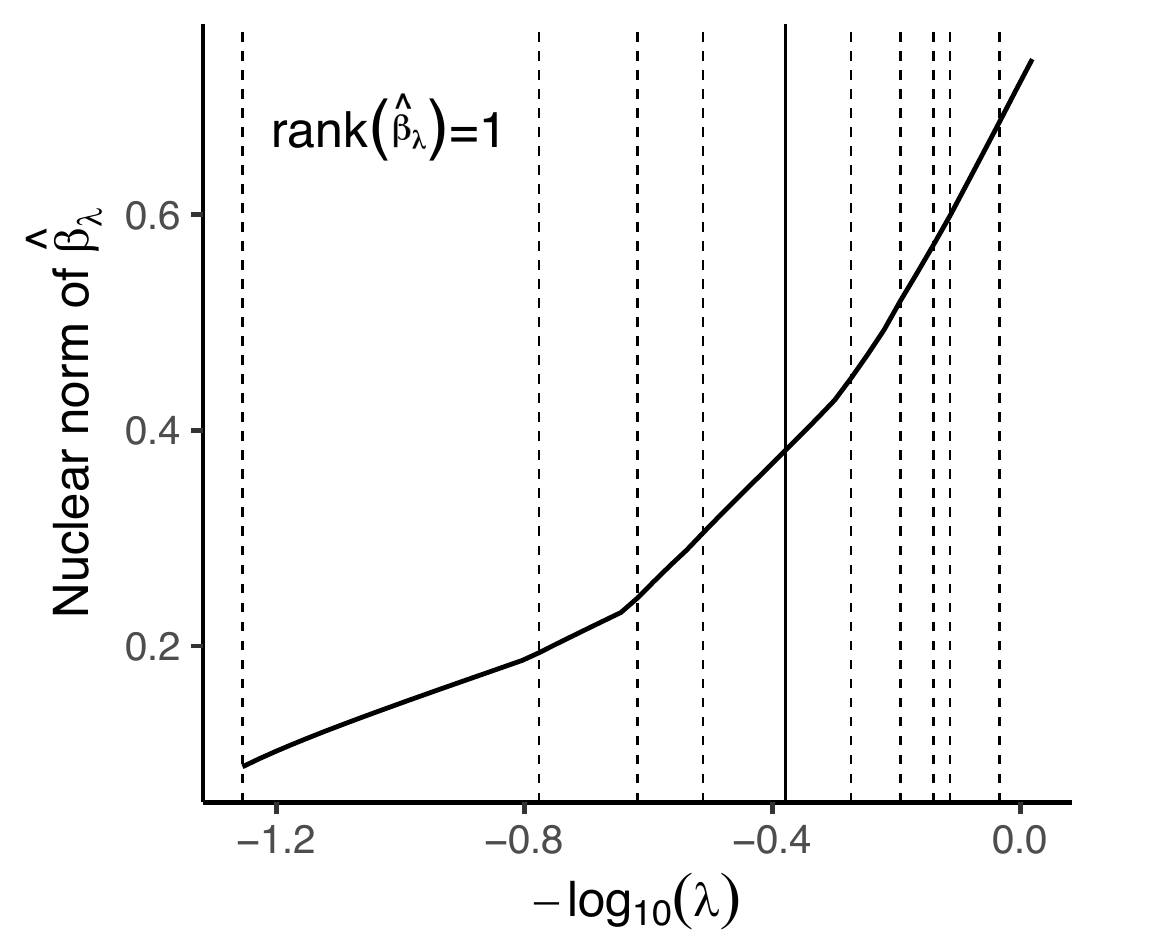} 
\includegraphics[width=11.0cm]{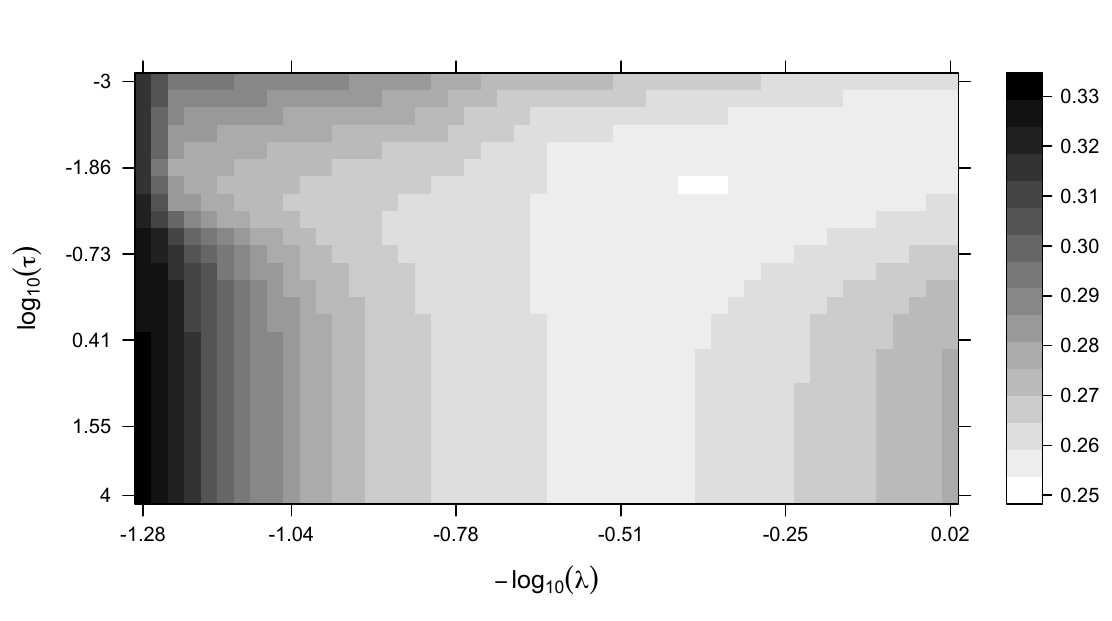}\hfill}
\centerline{\hfill{(a)}\quad\quad\quad\quad\quad\quad\quad\quad\quad\quad\quad\hfill\quad\quad\quad{(b)}\hfill}
\centerline{\hfill\includegraphics[width=18cm]{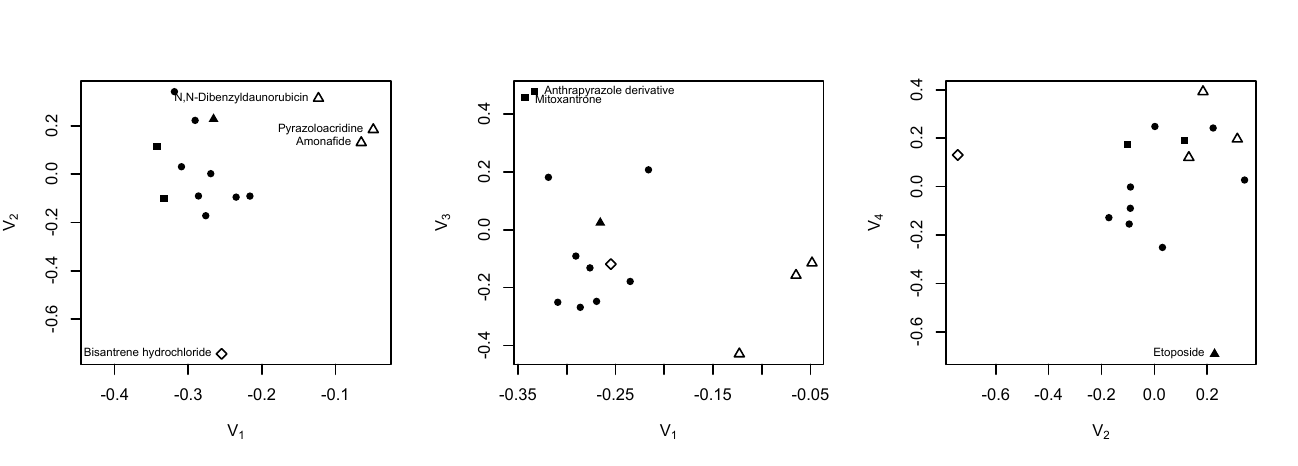}\hfill}
\centerline{\hfill{(c)}\hfill}
\caption{(a) A trace plot displaying the nuclear norm of $\widehat{\beta}_{\tau, \lambda}$ with $\tau = 0.0231$. Dashed vertical lines denote a change in rank, i.e., the dashed vertical line with $-\log_{10}(\lambda) \approx -1.5$ denotes the change from rank zero to rank one. The vertical dashed solid line denotes the value of $\lambda$ which minimizes the cross-validation squared prediction error. (b) A heatmap displaying the squared prediction error averaged across five folds and all 15 response for 25 candidate values of $\tau$ and 50 candidate values of $\lambda$. (c) Plots of the responses' factor loading values from the estimated regression coefficient matrix, i.e., columns of $V$ from the singular value decomposition $\widehat{\beta}_{\tau^*, \lambda^*} = UDV'$. Solid points denote drugs which are effective for treating leukemia; transparent triangles denote those effective for brain and spinal cancer and leukemia in mice; and the transparent diamonds denote those for effective for breast cancer. Note that symbols for each drug are the same across the three plots.}\label{fig:diagnostics} 
\end{figure}


In the analysis of \citet{cruz2014fast}, the authors used regularized canonical correlation analysis (CCA) on this particular dataset to examine low dimensional linear combinations of both microRNA expression and drug activity levels. Since CCA and reduced-rank regression are closely related, we analyzed this dataset using our method with a nuclear norm penalty \citep{yuan2007dimension}: see the bottommost row of Table 1 in the Supplementary Material for computational details and references.

First, we performed five-fold cross-validation using the entire dataset ($n=60$) to select tuning parameters. In Figure \ref{fig:diagnostics}(b), we display a heatmap of the squared prediction error averaged over all 15 drug responses and five folds. Notably, the $\tau$ selected by cross-validation is relatively small ($\tau = 0.0231$). For reference, when using nuclear norm-penalized least squares, the cross-validation squared prediction error is effectively equivalent to our method when $\tau = 10^4$ (i.e., the bottom row of Figure \ref{fig:diagnostics}(b)). In Figure \ref{fig:diagnostics}(a), we display the nuclear norm of the estimated regression coefficient matrix as a function the tuning parameter $\lambda$ with $\tau = 0.0231$ fixed. Dashed vertical lines indicate tuning parameter values where the estimated rank increases. The solid vertical line indicates the tuning parameter value which minimized squared prediction error averaged over the five folds and 15 responses. Let $(\tau^*, \lambda^*)$ denote the tuning parameter pair which minimized cross-validation squared prediction error. From this plot, we see that the estimated rank of the regression coefficient matrix was four in these data. When fitting the model using nuclear norm-penalized least squares, the estimated rank is only three. 

To interpret our estimated regression coefficient matrix $\widehat{\beta}_{\tau^*, \lambda^*}$, we examine the response factor loadings for the four factors. That is, letting $\widehat{\beta}_{\tau^*, \lambda^*} = UDV'$ be the singular value decomposition of $\widehat{\beta}_{\tau^*, \lambda^*}$, we plot the columns of $V$ in Figure \ref{fig:diagnostics}(c). These can be interpreted as response factor loadings since $XUD \in \mathbb{R}^{n \times 4}$ can be interpreted as the low-dimensional microRNA expression factors, so that $V \in \mathbb{R}^{q \times 4}$ can be interpreted as their loadings (i.e., $V'$ is the regression coefficient matrix for the predictors $XUD$). In Figure \ref{fig:diagnostics}(c), we display three two-dimensional pairs of the response factor loadings $V_k \in \mathbb{R}^{15}, (k=1, \dots, 4)$. In these plots, responses represented by solid points denote drugs which are effective for treating leukemia in humans, whereas transparent points are effective for other types of cancer or leukemia in mice. 

Focusing on the leftmost panel of Figure \ref{fig:diagnostics}(c), we see that the first factor loading ($V_1$) separates three responses (N,N-Dibenzyldaunorubicin, Pyrazoloacridine, and Amonafide) from the rest: these three drugs are effective for treating leukemia in mice and brain cancer (see Table 3 of \citet{cruz2014fast}). In this same plot, we see that loading two ($V_2$) separates the drug effective for breast cancer 
(Bisantrene hydrochloride) from the rest. Based on the two rightmost plots, it seems factors three and four separate two (Anthrapyrazole derivative and Mitoxantrone) and one (Etoposide), respectively, of the leukemia drugs from all other drugs. These findings are mostly consistent with those in \citet{cruz2014fast} who performed CCA on these data. In Figure 3 of the Supplementary Material, we display two-dimensional plots of the $XU$, which show that cell lines cluster according to their cancer type. 

\begin{table}[ht]
\centering
\scalebox{.80}{
\begin{tabular}{|r|cccc|cccc|}
  \hline 
 \multirow{2}{*}{Drug} & \multicolumn{4}{c|}{Average MSPE} & \multicolumn{4}{c|}{Median MSPE}\\
 & \texttt{NN-MC} & \texttt{NN-LS} & \texttt{Ridge} & \texttt{Null} & \texttt{NN-MC} & \texttt{NN-LS} &  \texttt{Ridge} & \texttt{Null} \\ 
  \hline
Doxorubicin & 27.60 & \cellcolor{gray!25}27.59 & 33.35 & 34.07 & \cellcolor{gray!25}16.42 & 17.41 & 24.41 & 21.24 \\ 
  Amonafide & \cellcolor{gray!25}4.29 & 4.45 & 5.11 & 4.60 & 3.54 & 3.51 & 3.79 & \cellcolor{gray!25}3.42 \\ 
  M-AMSA & \cellcolor{gray!25}34.93 & 35.70 & 42.23 & 51.61 & \cellcolor{gray!25}33.98 & 34.89 & 38.16 & 50.62 \\ 
  Anthrapyrazole derivative & \cellcolor{gray!25}34.06 & 34.79 & 40.81 & 47.81 & \cellcolor{gray!25}29.68 & 30.49 & 36.69 & 42.51 \\ 
  Pyrazoloacridine & \cellcolor{gray!25}7.90 & 8.25 & 9.74 & 8.40 & \cellcolor{gray!25}7.21 & 7.48 & 9.11 & 7.45 \\ 
  Bisantrene hydrochloride & 41.69 & \cellcolor{gray!25}41.07 & 46.72 & 42.99 & \cellcolor{gray!25}17.51 & 17.64 & 21.64 & 19.98 \\ 
  Daunorubicin & \cellcolor{gray!25}26.37 & 26.38 & 34.00 & 35.11 & \cellcolor{gray!25}20.13 & 20.98 & 29.09 & 31.25 \\ 
  Deoxydoxorubicin & \cellcolor{gray!25}27.36 & 27.37 & 33.53 & 33.21 & \cellcolor{gray!25}20.68 & 20.80 & 26.70 & 20.90 \\ 
  Mitoxantrone & \cellcolor{gray!25}38.44 & 39.36 & 48.61 & 52.60 & \cellcolor{gray!25}30.63 & 32.72 & 45.52 & 49.26 \\ 
  Menogaril & \cellcolor{gray!25}33.06 & 33.47 & 39.24 & 41.31 & \cellcolor{gray!25}29.92 & 30.64 & 34.76 & 34.59 \\ 
  N,N-Dibenzyldaunorubicin & \cellcolor{gray!25}19.18 & 20.04 & 21.30 & 22.87 & \cellcolor{gray!25}18.27 & 18.62 & 18.54 & 18.91 \\ 
  Oxanthrazole & \cellcolor{gray!25}14.94 & 15.05 & 17.92 & 20.29 & \cellcolor{gray!25}12.84 & 12.89 & 15.83 & 17.78 \\ 
  Rubidazone & \cellcolor{gray!25}20.47 & 20.51 & 24.21 & 25.06 & \cellcolor{gray!25}14.68 & 14.68 & 16.65 & 17.55 \\ 
  Teniposide & \cellcolor{gray!25}34.52 & 34.91 & 44.15 & 46.56 & \cellcolor{gray!25}27.01 & 27.06 & 34.42 & 32.55 \\ 
  Etoposide & 32.30 &\cellcolor{gray!25}31.57 & 38.02 & 44.76 & 31.80 & \cellcolor{gray!25}31.23 & 35.25 & 41.81 \\ 
   \hline
\end{tabular}
}
\caption{Average and median (over 500 independent training testing splits) mean squared prediction errors ($\times$100) for each of the fifteen drugs in the NCI-60 dataset we analyzed. Methods considered were the nuclear norm least squares estimator (\texttt{NN-LS}), the nuclear norm penalized version of \eqref{non_convex_estimator} (\texttt{NN-MC}), fifteen separate ridge regressions (\texttt{Ridge}), and the null model, i.e., the model assuming microRNA expression does not affect mean drug activity. Cells highlighted in gray are those with the lowest MSPE amongst the considered methods.}\label{table:drug_activity}
\end{table}

To verify that our estimator also provides an improvement in prediction accuracy, we performed additional cross-validation. For five hundred independent replications, we randomly selected five cell lines to be testing cell lines and fit the model using nuclear norm-penalized least squares, \eqref{non_convex_estimator} with a nuclear norm penalty, and separate ridge regressions. Tuning parameters for all methods were selected by five-fold cross-validation on the training data in each split. In Table \ref{table:drug_activity}, we display the average and median mean squared prediction error (over the five testing cell lines) of our method compared to nuclear norm penalized least squares, separate ridge regressions, and the null model. We observe that in all but one drug, our method provides a substantial improvement in prediction accuracy over the null model and separate ridge regressions. In the majority of drugs, our method outperforms the nuclear norm penalized least squares estimator. Notably, the estimates from our method had average rank 4.79, whereas the penalized least squares estimator had average rank 2.90. 
\section{Discussion}

We have proposed and studied a particular parametric link between the mean and error covariance in the multivariate response linear regression model. There are multiple important directions for future research.
  \begin{enumerate}[label=(\alph*),leftmargin=*]
  \item There are many further extensions to the regression model we
    consider.  For instance, as a referee pointed out, the decomposition of
    $\beta_*$ into a low-rank matrix plus a sparse matrix is ubiquitous, and
    thus, it would be useful to be able to apply our method in such settings.

  \item We have proposed a particular parametric link between the regression function and the error covariance matrix.
    An alternative type of link is assumed in envelope modelling \citep{cook2015foundations}.
    But there may be other link(s) that prove useful, and this could be a fruitful direction for future research to explore.

  \item Our method is based on a sum-of-squares (Frobenius) criterion for estimating the regression coefficients.  However, heavy-tailed and contaminated data are often encountered in multivariate response regression applications.  In such cases our criterion will not be effective. It would be useful to have alternatives for the Frobenius criterion, such as a Huberized-loss function, or an $L_1$ criterion function (which would connect the problem to median/quantile regression).  One difficulty in those settings is that the loss function, even without the added penalty, will be either nonconvex or nondifferentiable, so a new algorithm for computing the minimizer will need to be developed.
    
  \end{enumerate}

\vspace{-10pt}
\section*{Acknowledgments}
A.\ J.\ Rothman's research was supported in part by the National Science Foundation DMS-1452068. C.\ R.\ Doss's research was supported in part by the National Science
Foundation grants DMS-1712664 and DMS-1712706. The authors thank Dr. Raul Cruz-Cano for providing information about the NCI-60 dataset.

\section*{Supplementary Material}
Additional information, tables, and figures referenced in Sections 2-6 are available in the Supplementary Material online. An R package implementing our method is available for download at github.com/ajmolstad/MCMVR.

%
\linespread{1.5}\selectfont
\bibliography{arxiv_R1}

\begin{thebibliography}{}

\bibitem[Breiman and Friedman, 1997]{breiman1997predicting}
Breiman, L. and Friedman, J.~H. (1997).
\newblock Predicting multivariate responses in multiple linear regression.
\newblock {\em J. Roy. Statist. Soc. Ser. B}, 59(1):3--54.
\newblock With discussion and a reply by the authors.

\bibitem[Chen and Huang, 2012]{chen2012sparse}
Chen, L. and Huang, J.~Z. (2012).
\newblock Sparse reduced-rank regression for simultaneous dimension reduction
  and variable selection.
\newblock {\em J. Amer. Statist. Assoc.}, 107(500):1533--1545.

\bibitem[Chen and Sun, 2017]{chen2017prediction}
Chen, T.-H. and Sun, W. (2017).
\newblock Prediction of cancer drug sensitivity using high-dimensional omic
  features.
\newblock {\em Biostatistics}, 18(1):1--14.

\bibitem[Cook and Zhang, 2015]{cook2015foundations}
Cook, R.~D. and Zhang, X. (2015).
\newblock Foundations for envelope models and methods.
\newblock {\em J. Amer. Statist. Assoc.}, 110(510):599--611.

\bibitem[Cruz-Cano and Lee, 2014]{cruz2014fast}
Cruz-Cano, R. and Lee, M.-L.~T. (2014).
\newblock Fast regularized canonical correlation analysis.
\newblock {\em Computational Statistics \& Data Analysis}, 70:88--100.

\bibitem[Datta and Zou, 2017]{datta2017cocolasso}
Datta, A. and Zou, H. (2017).
\newblock Co{C}o{L}asso for high-dimensional error-in-variables regression.
\newblock {\em Ann. Statist.}, 45(6):2400--2426.

\bibitem[Gleser and Watson, 1973]{gleser1973estimation}
Gleser, L.~J. and Watson, G.~S. (1973).
\newblock Estimation of a linear transformation.
\newblock {\em Biometrika}, 60:525--534.

\bibitem[Huber, 1964]{huber1964}
Huber, P.~J. (1964).
\newblock Robust estimation of a location parameter.
\newblock {\em Ann. Math. Statist.}, 35(1):73--101.

\bibitem[Izenman, 1975]{izenman1975reduced}
Izenman, A.~J. (1975).
\newblock Reduced-rank regression for the multivariate linear model.
\newblock {\em J. Multivariate Anal.}, 5:248--264.

\bibitem[Lange, 2016]{lange2016mm}
Lange, K. (2016).
\newblock {\em M{M} Optimization Algorithms}.
\newblock Society for Industrial and Applied Mathematics, Philadelphia, PA.

\bibitem[Lee and Liu, 2012]{lee2012simultaneous}
Lee, W. and Liu, Y. (2012).
\newblock Simultaneous multiple response regression and inverse covariance
  matrix estimation via penalized {G}aussian maximum likelihood.
\newblock {\em J. Multivariate Anal.}, 111:241--255.

\bibitem[Li and Lin, 2015]{LiAcc2015}
Li, H. and Lin, Z. (2015).
\newblock Accelerated proximal gradient methods for nonconvex programming.
\newblock In Cortes, C., Lawrence, N.~D., Lee, D.~D., Sugiyama, M., and
  Garnett, R., editors, {\em Advances in Neural Information Processing Systems
  28}, pages 379--387. Curran Associates, Inc.

\bibitem[Liu et~al., 2010]{liu2010mrna}
Liu, H., D'Andrade, P., Fulmer-Smentek, S., Lorenzi, P., Kohn, K.~W.,
  Weinstein, J.~N., Pommier, Y., and Reinhold, W.~C. (2010).
\newblock {mRNA} and {microRNA} expression profiles of the {NCI-60} integrated
  with drug activities.
\newblock {\em Molecular Cancer Therapeutics}, 9(5):1080--1091.

\bibitem[Molstad and Rothman, 2016]{molstad2016indirect}
Molstad, A.~J. and Rothman, A.~J. (2016).
\newblock Indirect multivariate response linear regression.
\newblock {\em Biometrika}, 103(3):595--607.

\bibitem[Obozinski et~al., 2011]{obozinski2011support}
Obozinski, G., Wainwright, M.~J., and Jordan, M.~I. (2011).
\newblock Support union recovery in high-dimensional multivariate regression.
\newblock {\em Ann. Statist.}, 39(1):1--47.

\bibitem[Parikh et~al., 2014]{parikh2014proximal}
Parikh, N., Boyd, S., et~al. (2014).
\newblock Proximal algorithms.
\newblock {\em Foundations and Trends in Optimization}, 1(3):127--239.

\bibitem[Peng et~al., 2010]{peng2010regularized}
Peng, J., Zhu, J., Bergamaschi, A., Han, W., Noh, D.-Y., Pollack, J.~R., and
  Wang, P. (2010).
\newblock Regularized multivariate regression for identifying master predictors
  with application to integrative genomics study of breast cancer.
\newblock {\em Ann. Appl. Stat.}, 4(1):53--77.

\bibitem[Peng and Croce, 2016]{peng2016role}
Peng, Y. and Croce, C.~M. (2016).
\newblock The role of {MicroRNAs} in human cancer.
\newblock {\em Signal Transduction and Targeted Therapy}, 1:15004.

\bibitem[Perrot-Dock\`es et~al., 2018]{perrot2018variable}
Perrot-Dock\`es, M., L\'evy-Leduc, C., Sansonnet, L., and Chiquet, J. (2018).
\newblock Variable selection in multivariate linear models with
  high-dimensional covariance matrix estimation.
\newblock {\em J. Multivariate Anal.}, 166:78--97.

\bibitem[Pourahmadi, 1999]{pourahmadi1999joint}
Pourahmadi, M. (1999).
\newblock Joint mean-covariance models with applications to longitudinal data:
  Unconstrained parameterisation.
\newblock {\em Biometrika}, 86(3):677--690.

\bibitem[Pourahmadi, 2013]{pourahmadi2013high}
Pourahmadi, M. (2013).
\newblock {\em High-Dimensional Covariance Estimation: With High-Dimensional
  Data}.
\newblock John Wiley \& Sons.

\bibitem[Rothman et~al., 2010]{rothman2010sparse}
Rothman, A.~J., Levina, E., and Zhu, J. (2010).
\newblock Sparse multivariate regression with covariance estimation.
\newblock {\em J. Comput. Graph. Statist.}, 19(4):947--962.

\bibitem[Shoemaker, 2006]{shoemaker2006nci60}
Shoemaker, R.~H. (2006).
\newblock The {NCI-60} human tumour cell line anticancer drug screen.
\newblock {\em Nature Reviews Cancer}, 6(10):813.

\bibitem[Velu and Reinsel, 2013]{velu2013multivariate}
Velu, R. and Reinsel, G.~C. (2013).
\newblock {\em Multivariate reduced-rank regression: theory and applications},
  volume 136.
\newblock Springer Science \& Business Media.

\bibitem[Yin and Li, 2011]{yin2011sparse}
Yin, J. and Li, H. (2011).
\newblock A sparse conditional {G}aussian graphical model for analysis of
  genetical genomics data.
\newblock {\em Ann. Appl. Stat.}, 5(4):2630--2650.

\bibitem[Yuan et~al., 2007]{yuan2007dimension}
Yuan, M., Ekici, A., Lu, Z., and Monteiro, R. (2007).
\newblock Dimension reduction and coefficient estimation in multivariate linear
  regression.
\newblock {\em J. R. Stat. Soc. Ser. B Stat. Methodol.}, 69(3):329--346.

\end{thebibliography}
\newpage





\end{document}